\documentclass[aps,pra,superscriptaddress,12pt,tightenlines,nofootinbib]{revtex4}
\usepackage{amsmath,amsthm,graphicx,amssymb,verbatim,listings}
\usepackage{wasysym}
\usepackage[T1]{fontenc}     % needed for the guillemets
\usepackage{lmodern}         % needed to fix suboptimal font rendering
\usepackage[ pdftex, plainpages = false, pdfpagelabels,
                 pdfpagelayout = useoutlines,
                 bookmarks,
                 bookmarksopen = true,
                 bookmarksnumbered = true,
                 breaklinks = true,
                 linktocpage=all,
                 pagebackref=false,
                 colorlinks = true,
                 linkcolor = BrickRed,
                 urlcolor  = blue,
                 citecolor = BrickRed,
                 anchorcolor = green,
                 hyperindex = true,
                 hyperfigures
                 ]{hyperref}
\usepackage[usenames, dvipsnames]{xcolor}
\usepackage{xifthen} % provides \isempty test

\newcommand{\tr}{\hbox{tr}}

\newcommand{\ket}[1]{{\ensuremath{\left| #1 \right\rangle}}}
\newcommand{\bra}[1]{{\ensuremath{\left\langle #1 \right|}}}
\newcommand{\braket}[2]{{\ensuremath{\left\langle #1 \middle| #2
      \right\rangle}}}
\newcommand{\ketbra}[2]{{\ensuremath{\left| #1 \middle\rangle\!\middle\langle #2
      \right|}}}

\newcommand{\myurl}[2][]{\ifthenelse{\isempty{#1}}{\url{#2}}{\href{#1}{\tt #2}}}
\newcommand{\quantph}[1]{\href{http://arxiv.org/abs/quant-ph/#1}{{\tt quant-\allowbreak{}ph/\allowbreak{}#1}}}
\newcommand{\arxiv}[2][]{\ifthenelse{\isempty{#1}}{\href{http://arxiv.org/abs/#2}{{\tt arXiv:\allowbreak{}#2}}} {\href{http://arxiv.org/abs/#2}{{\tt arXiv:\allowbreak{}#2 [#1]}}}}
\newcommand{\pirsa}[1]{\href{http://pirsa.org/#1/}{{\tt PIRSA:\allowbreak{}#1}}}
\newcommand{\booktitle}{\textsl}
\newcommand{\hrefdoi}[2]{\href{https://dx.doi.org/#1}{#2}}

\newcommand{\cP}{\mathcal{P}}

\newcommand{\cH}{\mathcal{H}}

\newcommand{\bbR}{\mathbb{R}}
\newcommand{\bbC}{\mathbb{C}}

\newtheorem{bub}{Bubism}
\newcommand{\bjb}{\begin{bub}\protect$\!\!${\em }$\;\;$}
\newcommand{\ejb}{\end{bub}}

\newtheorem{fraassen}{van Fraassenism}
\newcommand{\bvf}{\begin{fraassen}\protect$\!\!${\em }$\;\;$}
\newcommand{\evf}{\end{fraassen}}

\newcommand{\bq}{\begin{quotation}}
\newcommand{\eq}{\end{quotation}}

\begin{document}

\title{Are Non-Boolean Event Structures the Precedence or Consequence of Quantum Probability?}

\author{Christopher A. Fuchs}
\author{Blake C. Stacey}
\affiliation{\href{http://www.physics.umb.edu/Research/QBism/}{Physics Department}, University of Massachusetts Boston}

\date{\today}

\begin{abstract}
  In the last five years of his life Itamar Pitowsky developed the
  idea that the formal structure of quantum theory should be thought
  of as a Bayesian probability theory adapted to the empirical
  situation that Nature's events just so happen to conform to a
  non-Boolean algebra.  QBism too takes a Bayesian stance on the
  probabilities of quantum theory, but its probabilities are the
  personal degrees of belief a sufficiently-schooled agent holds for
  the consequences of her actions on the
  external world.  Thus QBism has two levels of the personal where the
  Pitowskyan view has one.  The differences go further.  Most
  important for the technical side of both views is the quantum
  mechanical Born Rule, but in the Pitowskyan development it is a
  theorem, not a postulate, arising in the way of Gleason from
  the primary empirical assumption of a non-Boolean algebra.  QBism on
  the other hand strives to develop a way to think of the Born Rule in
  a pre-algebraic setting, so that it itself may be taken as the
  primary empirical statement of the theory.  In other words, the hope
  in QBism is that, suitably understood, the Born Rule is quantum
  theory's most fundamental postulate, with the Hilbert space
  formalism (along with its perceived connection to a
  non-Boolean event structure) arising only secondarily.  This paper
  will avail of Pitowsky's program, along with its extensions in the
  work of Jeffrey Bub and William Demopoulos, to better explicate
  QBism's aims and goals.
\end{abstract}
\maketitle

\section{Dedication by CAF}

This paper was written too late to appear in the memorial volume for
Itamar Pitowsky edited by Meir Hemmo and Orly
Shenker~\cite{HemmoShenker:2020}, but it is never too late to tribute
dear friends.  Itamar Pitowsky and Bill Demopoulos came into my life
as a package.  I met Itamar in 1997 at the Sixth UK Conference on
Conceptual and Mathematical Foundations of Modern Physics in Hull,
England. It was my first quantum foundations conference, and to tell
the truth I was a little ashamed to be there: Indeed I had to work up
the nerve to ask my postdoctoral advisor John Preskill if I might use
some of my funding for such a frilly thing.  And it was a frilly
thing --- only one talk in the whole meeting made any impression on me,
Itamar's on simplifying Gleason's theorem. It blew me away.  When I
got home, I studied his paper~\cite{Pitowsky:1997} with meticulous
care, not just for the math, but also for its programmatic character.
I found something so alluring in how he presented the big picture of
his efforts at the end, I pledged to one day write a paper myself
ending with nearly the same words: ``More broadly, Theorem 6 is part
of the attempt to understand the mathematical foundations of quantum
mechanics. In particular, it helps to make the distinction between its
physical content and mathematical artifact clear.''

Well, I must have made an impression on Itamar too, as he described me
to Bill Demopoulos as ``a Wunderkind of quantum information''. (If we
could only recapture our youths!)  One thing led to another, from more
than 1,300 email exchanges, to many long, thoughtful walks at Bill's
farm in Ontario.  Both men made an indelible impression on
me. Professionally, what set them apart from other philosophers of
physics is that, rather than try to dissuade me from developing QBism,
they both were sources of constant encouragement.  But there was so
much more than our professional relationships between us.  At least in the case
of Bill's passing, it was not so sudden as Itamar's that I was able to
tell him I love him.

This paper is dedicated to Itamar and Bill.

\section{Introduction}

QBism is a relative newcomer among interpretations of quantum
mechanics.  Though it has identifiable precursors, no mature statement
of it existed in the literature before 2009--2010~\cite{Fuchs:2009,
  Fuchs:2010}.  Indeed, even the pre-QBist writings of the developers
themselves contain much that QBism flatly
contradicts~\cite{Stacey:2019c}.  One brief summary is that QBism is
\begin{quotation}
\noindent an interpretation of quantum mechanics in which the ideas of
\emph{agent} and \emph{experience} are fundamental. A ``quantum
measurement'' is an act that an agent performs on the external
world. A ``quantum state'' is an agent's encoding of her own personal
expectations for what she might experience as a consequence of her
actions. Moreover, each measurement outcome is a personal event, an
experience specific to the agent who incites it. Subjective judgments
thus comprise much of the quantum machinery, but the formalism of the
theory establishes the standard to which agents should strive to hold
their expectations, and that standard for the relations among beliefs
is as objective as any other physical theory~\cite{DeBrota:2018}.
\end{quotation}
This is not the kind of move that one takes lightly.  Only the
stunning empirical success of quantum mechanics, coupled with the
decades of confusion over its conceptual foundations, could have
compelled the codification of thoughts like these into a detailed,
systematized research program.  The practical triumphs of quantum
physics imply that the questions of conceptual foundations should be
taken seriously, while the perennial debate, ceaseless and without
clarification, implies that a radical step is necessary.

When compared against most interpretations of the quantum, QBism
requires perspective shifts of an almost virtiginous character.  For
starters, it insists upon a stringently personalist Bayesianism.  The
problem with physicists and probability is that when one has vast
heaps of data, pretty much any even half-baked interpretation of
probability is good enough to scrape by.  It is not until we push
further into matters of principle that mediocrity begins to pose a
real hazard.  Fortunately, physicists are trained to shed intuitions
that prove counterproductive.  We could once get by with acting like
the data points which help shape probability assignments actually
\emph{determine} them, but then, we also got along for quite a while
thinking that heavy bodies must fall faster than light ones.

Likewise, foregrounding the concept of \emph{agent} means taking a
first-person perspective, and first-person singular at
that~\cite{Mermin:2014}.  This means rephrasing, and sometimes even
rethinking, some discussions of quantum matters.  For example, a
first-person take on quantum communication protocols is that,
fundamentally, Alice wants to be confident that Bob will react in the
way she needs after she has intervened in the world by sending a
message.  For if Alice had no concern for the consequences returning
to her or the potential suffering of those she regards as kin, then
she would abandon the photonics and just throw a bottle into the sea.

Some changes in language end up being rather modest, though, in scale
if not in eventual implication.  QBism makes no category distinction
between quantum states and probability distributions; the former are
just examples of the latter in a coordinate system adapted for atomic
and molecular physics~\cite{Fuchs:2019}.  Thus, the QBist avoids
locutions like ``the quantum state of the system'' in favor of
``\emph{my} quantum state \emph{for} the system'' --- not too many
letters changed, but with a great deal of history behind the
difference.  A QBist exposition might say, ``Suppose a physicist Alice
ascribes a quantum state $\rho$,'' a turn of phrase that is
philosophically justified while also sounding quite natural, since
ascribing quantum states is a thing that working physicists actually
do.

This example illustrates the peculiar flipside of QBist radicalism.
After traversing the wilds of personalist Bayesian probability and the
further realms of participatory realism~\cite{Fuchs:2016a}, one
arrives with a new appreciation of home.  Certain habits of more
innocent days turn out to be philosophically justified, or at least
justifiable; one learns that one can do quantum foundations and
quantum \emph{physics} at the same time.

So, now that we've gotten a little bit of seeming almost-paradox on our
fingers, it would be good to sort out what is and is not a shocking
departure.  How does QBism situate itself with respect to those other
interpretational programs that share some of its morphology, either by
convergence or common ancestry?

\section{Two Levels of Personalism}

Introducing a collection of correspondence, CAF identified
three characteristics of QBism that distinguish it
from prior interpretations~\cite[p.\ ix]{Fuchs:2014}.
\begin{quotation}
  \noindent First is its crucial reliance on the mathematical tools of
  quantum information theory to reshape the look and feel of quantum
  theory's formal structure. Second is its stance that two levels of
  radical ``personalism'' are required to break the interpretational
  conundrums plaguing the theory. Third is its recognition that with
  the solution of the theory's conundrums, quantum theory does not
  reach an end, but is the start of a great journey.
\end{quotation}
We will return to the first point later (it is the motivation for the
technical work that occupies most of our time not given over to
teaching and administration).  In this section, we will focus on the
second point, the need for two levels of personalism.  A pivotal
moment in the history of QBism was the realization that one level did
not suffice, even though that was already one more than many were
willing to grant!

We have already met a significant part of the first layer in the
introduction, where we noted that in QBism, quantum states are the
possessions of the agent who asserts them.  This is already enough to
get booted from a philosophy-of-physics meeting, but we've hardly
gotten started.  Chasing the demands of self-consistency, we find that
other elements of the quantum formalism must \emph{also} be given the
same status as quantum states, that is, the status of personalist
Bayesian probability assignments.  One important example is \emph{the
  association of specific POVM elements with potential future
  experiences.}  A positive-operator-valued measure, or POVM for
short, is the basic notion of measurement in quantum information
theory.  Briefly put, a POVM is a collection of positive semidefinite
operators on a Hilbert space that sum to the identity operator on that
space, and each of the operators (or ``effects'') in the set
corresponds to a possible outcome of the measurement that the POVM
represents~\cite{Peres:1993, Nielsen:2010}.

In contrast, in the tradition of Bub, Demopoulos, and Pitowsky one
moves from quantum to classical by replacing the space of possible
events with a non-Boolean algebra.  Quantum probability is Bayesian
probability on this algebraic structure.  Events are tied to
projectors on Hilbert space, and this binding is in essence an
objective fact of nature~\cite{Bub:2007}.  This leads quickly to
trouble~\cite{Fuchs:2002}.  For suppose that a physicist Alice starts
her day with a subjective probability assignment, encoded as a density
operator $\rho$.  She then performs a measurement and obtains an
outcome, which will be associated with some projector $\Pi$.  She
updates her state assignment accordingly, replacing $\rho$ with~$\Pi$.
But now, \emph{her probabilities have been locked to an objective
  value.}  In order for the personalism of quantum-state assignments
not to collapse into triviality, there has to be more flexibility.
(For additional discussion on this point, see the
\hyperref[10january2007]{10 January 2007} entry drawn from
correspondence with Bub, Demopoulos, Pitowsky, and Veiko Palge in
Appendix B.)

The thorn that spurred the development of the second level of
personalism was the old thought-experiment known as ``Wigner's
Friend'' or the puzzle of the observer being
observed~\cite{Fuchs:2019}.  This is a tale --- \emph{not} as old as
time, though within the insular world of quantum foundations, it may
seem so --- a tale of two agents and one quantum system.  The agents
could be Wigner and his friend, or in the lingo of quantum
information, Alice and Bob.  The story begins with Alice and Bob
having a conversation.  They both agree, let us say, that a quantum
state $\ket{\psi}$ expresses their beliefs about the system.  (If they
are inclined to the Bayesian school, they don't \emph{have} to agree
at this stage.  $\ket{\psi}$ is a catalogue of expectations, or
bluntly put, a compendium of beliefs.  But let's say they do agree so
far.)  Moreover, they agree that at a given time, Alice will
perform an experiment upon the system, a measurement which both Alice
and Bob say is represented by some mathematical structure such that if
Alice observes outcome $i$, she will update her expectations to the
state vector $\ket{i}$.  Then, Bob walks away and leaves the room.
Alice is alone with the system.  Time passes, and the agreed-upon
moment of measurement goes by.

What, now, is the ``correct'' state that each physicist should have
chosen for the system?  Alice, everyone concurs, should have selected
some vector $\ket{i}$.  But what about Bob?  Unless he holds Alice
somehow exempt from quantum mechanics, he would express his beliefs
about her, in principle, with a quantum state $\rho$, and a
time-evolution operator family $U_t$ with which he can adjust those
expectations and calculate probabilities pertaining to different
times.  Before Bob interacts with Alice or the other system, he would
write a joint state for the two, a quantity of the form
\begin{equation}
  U_t \left(\rho \otimes \ketbra{\psi}{\psi}\right) U_t^\dag
\end{equation}
which will in general be an entangled state.  Bob's state for the
system will be a partial trace of this, typically a mixed state
completely different from $\ket{i}$.

Either Alice or Bob could be deemed ``incorrect''; for if either ``had
all the information'' they would arguably be forced to adopt the state
assignment of the other.
\begin{quotation}
  \noindent And so the back and forth goes.  Who has the {\it right\/}
  state of information?  The conundrums simply get too heavy if one
  tries to hold to an agent-independent notion of correctness for
  otherwise personalistic quantum states.  A QBist dispels these and
  similar difficulties by being conscientiously forthright.  {\it
    Whose information?\/} ``Mine!''  {\it Information about what?\/}
  ``The consequences (for {\it me\/}) of {\it my\/} actions upon the
  physical system!''  It's all ``I-I-me-me mine,'' as the Beatles
  sang~\cite{Fuchs:2019}.
\end{quotation}

Lately, there has been a resurgence of interest in Wigner's Friend ---
a growth industry, by the standards of quantum foundations.  These
recent elaborations of Wigner's Friend wrap it around a
no-hidden-variables argument~\cite{Frauchiger:2017, Brukner:2017,
  Pusey:2018, Bub:2019}.  QBism, having answered the original
conundrum on its own terms, finds nothing troubling about these
elaborations.  (Pusey regards an extended Wigner's Friend puzzle as
``actually good news'' for QBism and for Rovelli's Relational Quantum
Mechanics~\cite{Pusey:2018}. For a comparison and contrast between
QBism and Rovellian RQM, see~\cite{DeBrota:2018}; for a critical
evaluation of various statements made about QBism in this niche of the
literature, see~\cite{Stacey:2019}.)

\section{Gleason's Theorem}

Gleason's theorem is, in our experience, less well-known among
physicists than some of its conceptual descendants and
confr\`eres. The original statement had the mildly intimidating
language of finding ``all measures on the closed subspaces of a
Hilbert space''~\cite{Gleason:1957}.  In more detail, Gleason defined
a \emph{frame function} of weight $W$ for a separable Hilbert space
$\cH$ to be a function $f:\cH \to \bbR$ such that if $\{x_i\}$ is an
orthonormal basis of~$\cH$ then
\begin{equation}
  \sum f(x_i) = W.
\end{equation}
A frame function is \emph{regular} if there exists a self-adjoint
operator $T$ acting on $\cH$ such that $f(x)$ is given by an inner
product
\begin{equation}
  f(x) = (Tx, x)
\end{equation}
for all $x \in \cH$.  Gleason proves, by ingenious and somewhat
arduous means, that all nonnegative frame functions in three or more
dimensions are regular.

If the weight of a frame function is 1, then that function begins to
look like a probability assignment.  In physics language, Gleason
showed that if measurement outcomes are represented by vectors in
orthonormal bases, then any consistent ascription of probabilities to
measurement outcomes must take the form of the Born Rule. The
definition of frame functions assumes that probability ascriptions are
\emph{noncontextual}~\cite{Barnum:2000}: The function $f$ sends a
vector to a number in the unit interval, regardless of what basis that
vector may be a part of.  Rather remarkably, Gleason's proof yields
both the set of valid density operators and the Born Rule.  Quite an
economy of postulates!  It is easy to see why an attempt to rebuild
quantum theory on a better axiomatic footing might proceed by way of
Gleason's theorem.  As Wilce put it~\cite{Wilce:2017},
\begin{quotation}
  \noindent The point to bear in mind is that, once the
  quantum-logical skeleton [the lattice of closed subspaces of~$\cH$]
  is in place, the remaining statistical and dynamical apparatus of
  quantum mechanics is essentially fixed.  In this sense, then,
  quantum mechanics --- or, at any rate, its mathematical framework
  --- \emph{reduces to} quantum logic and its attendant probability
  theory.
\end{quotation}

Why, then, have the QBist and QBist-adjacent efforts at reconstructing
quantum theory skipped past Gleason's theorem?  Part of the answer
lies in that invocation of ``the quantum-logical skeleton'': As any
good biologist knows, bones are not simple affairs~\cite{Switek:2019},
and presuming a working skeleton is asking for quite the sophisticated
development already!

Gleason begins with a Hilbert space and orthonormal bases upon it.
How, then, should we arrive at Hilbert space?  Pitowsky presents an
argument for getting to the point where Gleason can take over, a
\emph{representation theorem} for lattices that satisfy a certain set
of conditions inspired by quantum mechanics~\cite{Pitowsky:2006}.  He
admits a gap in this argument, partly closed by another remarkable
theorem, this one by Maria Pia Sol\`er~\cite{Soler:1995,
  Prestel:1995}.  But the QBist concern is more fundamental: Those
lattice-theoretic conditions hail from \emph{the quantum physics of
  the 1930s,} when ``uncertainty'' was the shocking feature, long
before the discovery of more subtle mysteries like the violation of
Bell inequalities.  We can always take a fundamentally classical
theory, a theory of local degrees of freedom, and \emph{hide those
  variables} by imposing an ``uncertainty
principle''~\cite{Spekkens:2007, Spekkens:2016}.  The resulting
theories include features like the no-cloning theorem, incompatible
observables, and even interference between paths in a Mach--Zehnder
interferometer.  This modern perspective, informed by studies into
questions like what a quantum computer requires to outperform a
classical emulation of one, takes the glamor away from the notions in
which quantum logic was grounded.

The original version of Gleason's theorem fails if the Hilbert space
is two-dimensional. Why? When $d = 2$, one cannot hold one basis
vector in place and twirl the others around to generate multiple
distinct bases. So, the constraints assumed by Gleason are too weak to
carry real implicative weight. One might, for example, paint the
northern hemisphere of the Bloch sphere all over with 1's, and the
southern hemisphere with 0's.  The resulting assignment of
probabilities to von Neumann measurements --- remember, orthonormal
bases in $d = 2$ are pairs of antipodal points on the Bloch sphere ---
would be consistent with Gleason's rules, yet no choice of density
operator $\rho$ can yield those probabilities.  This pesky difficulty
is obviated if one follows the practice of quantum information theory
and broadens the notion of \emph{measurement} beyond the von Neumann
definition to the wider class of POVMs. Doing so allows a proof of a
Gleason-type result that is both simpler to prove and applicable in $d
= 2$ where Gleason's was not~\cite{Busch:2003, Caves:2004}. For
additional discussion, see~\cite{Granstroem:2006, Wright:2019a,
  Wright:2019b}.

The general good cheer created by shifting the fundamental notion of
measurement to POVMs led QBism further from the quantum-logic tradition.

\section{Non-Booleanity as a Consequence}
\label{sec:tech}
In this section, we will derive non-Booleanity as a consequence of a
more fundamental manifestation of quantum strangeness, the
nonexistence of intrinsic hidden variables.  We will step through this
argument for the case where it is simplest, a single qubit, which is
already enough to derive a theory with noncommuting observables.
Having gotten that far, we will briefly outline two approaches to
generalizing the argument and thus deriving quantum theory in higher
dimensions.

The centerpiece of the QBist program for reconstructing quantum theory
is the concept of a \emph{reference measurement.}  This is an
experiment with the property that if an agent possesses a probability
distribution over its outcomes, she can calculate the probabilities
that she should use for any other measurement.  In classical particle
mechanics, a reference measurement would be an experiment that reads
off the position and momentum of all particles in the system, thereby
locating it within phase space and allowing the agent to extrapolate
the full dynamical trajectory.  Any other experiment --- say,
observing the total angular momentum --- is essentially a
coarse-graining of the information that the reference measurement
provides.  If Alice expresses her uncertainty about the system's
phase-space coordinates as a probability distribution over phase
space, then she can calculate her probabilities for the possible
results of a coarse-grained observation by convolving that
``Liouville density'' with an appropriate kernel.

Classical mechanics has the property that if two Liouville densities
are distinguishable by a coarse-grained measurement, then they are
nonoverlapping.  Quantum physics violates this principle, in a way
that we can make precise.

Consider the following scenario~\cite{Appleby:2017}.  An agent Alice
has a physical system of interest that she plans to perform either one
of two experimental protocols upon.  In one protocol, she will send
the system directly into a measuring device and obtain an outcome.  In
the other, she will pass the system through a reference measurement
first and then send it into the device from the first protocol.  Let
$H_i$ denote the possible outcomes of the reference measurement and
$D_j$ the possible outcomes of the other.  Then Alice can write her
expectations for the results of these two protocols as $P(H_i, D_j)$,
for sending the system into the $D_j$ device via the reference $H_i$
experiment first, and $Q(D_j)$, for going directly into the $D_j$
device.  Probability theory itself imposes no bond between these
expectations: different conditions, different probabilities!  But the
additional assumption that the reference measurement simply reads off
the pre-existing physical condition of the system, \emph{an
  ontological assumption on top of probability theory,} does furnish a
link:
\begin{equation}
  Q(D_j) = \sum_i P(H_i, D_j)
  = \sum_i P (H_i) P(D_j | H_i).
\end{equation}
The first equality here is a \emph{physical} assumption, while the
second is mandated by logic.  That is, if Alice rejects the second
equality --- the ``Law of Total Probability'' --- she is being
Dutch-book incoherent, whereas if she rejects the first, she is just
being skeptical about the applicability of classical intuition.
Quantum physics provides a replacement for the first equality.
Instead of marginalizing over $H_i$ as in the previous expression, one
performs a ``quantum marginalization'':
\begin{equation}
  Q(D_j) = \mu\left(P(H_i, D_j)\right).
\end{equation}
The exact form of the function $\mu$ will depend upon the choice of
POVM used as the reference measurement.

To make this more concrete, let us take the case where the system of
interest is a single qubit.  Then a reference measurement can be any
POVM whose effects span the space of Hermitian operators on the
Hilbert space $\bbC^2$.  For example, let the reference measurement be
a POVM whose effects are proportional to rank-1 projectors that form a
regular tetrahedron inscribed in the Bloch sphere.  A convenient
indexing is to let $a$ and $b$ take the values $\pm 1$, and to write
\begin{equation}
  H_{ab} = \frac{1}{4}\left(
  I + \frac{1}{\sqrt{3}}(a\sigma_x + b\sigma_y + ab\sigma_z)
  \right).
  \label{eq:qubit-sic}
\end{equation}
These four operators sum to the identity and span the space as
desired.  With
\begin{equation}
  P(H_{ab}) = \tr(\rho H_{ab})
\end{equation}
by the Born Rule, we have
\begin{equation}
  Q(D_j) = \tr(\rho D_j)
  = \sum_{ab}\left[3 P(H_{ab}) - \frac{1}{2}\right]
  P(D_j|H_{ab}).
  \label{eq:qubit-urgleichung}
\end{equation}
Here, we have defined
\begin{equation}
  P(D_j|H_{ab}) = 2\tr(D_j H_{ab})
\end{equation}
as Alice's probability of obtaining the outcome $D_j$ \emph{given}
that her preparation for the system is the state proportional to~$H_{ab}$.

In this case, the function $\mu$ takes the form of the classical Law
of Total Probability but with an elementwise affine map applied to the
probability vector $P(H_{ab})$. Said another way,
Eq.~(\ref{eq:qubit-urgleichung}) \emph{is} the Born Rule, written in
explicitly probabilistic language throughout.

We see that picking a reference measurement establishes a mapping from
density matrices into a probability simplex, thereby furnishing a
wholly probabilistic way to write the quantum theory of a qubit.
Importantly, in this representation not all probability vectors
correspond to valid quantum states.  The state space $\cP$ is the ball
in~$\bbR^4$ within which
\begin{equation}
 \frac{1}{6} \leq \sum_{a,b} P(H_{ab}) P'(H_{ab})
\leq \frac{1}{3},\ \forall\ P,P' \in \cP.
\end{equation}
The pure states will be exactly those that saturate the upper bound on
their Euclidean norm.  Orthogonal states are \emph{maximally distant,}
saturating the lower bound, but the lower bound is not zero.  As we
prove in Appendix~\ref{app:ortho}, this is in fact a very general
result, holding true for any reference measurement: Quantum states
that are orthogonal in Hilbert space are overlapping probability
distributions.

We can turn this idea around.  Let's forget quantum theory for the
moment and take the relation
\begin{equation}
  Q(D_j) = \sum_{i=1}^N \left[\alpha P(H_i) - \beta\right] P(D_j|H_i)
  \label{eq:urgleichung}
\end{equation}
as the defining axiom of a probabilistic theory.  That is, let's say
that this expression, which ``breaks'' classical intuition in the
minimal way possible, is the empirical rule on top of probability
theory that tells us how to ``marginalize'' over an experiment that is
not merely ignored, but \emph{unperformed.}  Probability vectors
$P(H)$ and conditional probability matrices $P(D|H)$ are consistent
with one another if they always yield valid probability vectors $Q(D)$
when combined in this way.  This provides a reciprocal consistency
condition between the set of valid $P(H)$ and the set of valid
$P(D|H)$, which we can turn into a condition on the set of valid
$P(H)$ by making some fairly mild assumptions relating those two sets.
For instance, we can posit that the state of maximal indifference, the
flat probability vector, is a valid $P(H)$, and that posteriors from
maximal indifference are valid priors.  (For some conceptual
background, see the \hyperref[7january2011]{7 January 2011} entry
drawn from correspondence with Demopoulos in Appendix B.)  This
implies that, if the $P(H)$ are column vectors, then any row in a
matrix $P(D|H)$ is the transpose of some valid $P(H)$, up to overall
scaling:
\begin{equation}
  P(D_j|H_i) = \gamma P'(H_i).
\end{equation}
The condition that $Q(D_j)$ be always nonnegative then yields
\begin{equation}
  \alpha \sum_i P(H_i)P'(H_i) - \beta \sum_i P'(H_i) \geq 0,
\end{equation}
which by normalization simplifies to
\begin{equation}
  P \cdot P' \geq \frac{\beta}{\alpha}.
\end{equation}
Pairs that saturate this bound are maximally distant from one another
and imply a $Q(D_j)$ of 0. Moreover, because $Q(D)$, $P(H)$ and
$P(D|H)$ must themselves be properly normalized, we have that
\begin{equation}
  1 = \alpha - N\beta.
\end{equation}

We follow the guiding principle that our theory should be as broadly
applicable as possible, so the structures within it should have no
``distinguishing marks or scars'' that would make them peculiar to
particular physical situations.  Likewise, we want the sets we derive
to be \emph{maximal} with respect to the constraints we apply:
To do otherwise would be to implicitly admit an additional constraint,
and thus to be making more empirical assumptions on top of the basic
``nonclassical marginalization'' rule originally posited.  Therefore,
our set $\cP \subset \bbR^N$ of valid probability vectors must have
the property that any two points in it have a Euclidean inner product
greater than $\beta/\alpha$, and including \emph{any additional point}
would spoil that property.

This is already enough to make available some useful tools from
higher-dimensional geometry.  Given a point $P$ in the probability
simplex $\Delta$, the \emph{polar} of $P$ is the set of all vectors in
$\mathbb{R}^N$ whose elements sum to unity and which have an inner
product with $P$ that is greater than or equal to $\beta/\alpha$.  The
polar of a set is the set of all vectors that are in the polars of all
the points in the given set:
\begin{equation}
  A^* = \left\{ u : \sum_i u(i) = 1,
  \ u \cdot v \geq \frac{\beta}{\alpha}\ \forall v \in A
  \right\}.
\end{equation}
The requirement that a state space $\cP$ be maximal implies that it is
a \emph{self-polar} subset of the probability simplex~\cite{Appleby:2017}.

To move in the direction of quantum theory, we introduce a concept of
dimension in terms of \emph{Mutually Maximally Distant} sets.  For any
self-polar subset of the probability simplex, there must be a smallest
sphere that encloses it, i.e., the circumscribing sphere of the state
space.  The states on this sphere are states of maximal confidence
with respect to the reference measurement.  We define a Mutually
Maximally Distant set in $\cP$ to be a set of points on the
intersection of $\cP$ and its circumscribing sphere such that any two
points in the set saturate the lower bound on the inner product,
$\beta/\alpha$.  This is the tool we need to establish an
information-theoretic notion of ``dimension'': The dimension of a
system is the largest possible size of a set of hypotheses, each of
which are assertions of maximal confidence, and any two of which are
maximally distinct from each other.  (This is the natural counterpart
of classical probability theory, where hypotheses of maximal
confidence are Kronecker delta functions.)  A brief
derivation~\cite{Appleby:2017, Appleby:2011} shows that
\begin{equation}
  d = 1 + \frac{r_{\rm out}^2}{r_{\rm mid}^2},
\end{equation}
where $r_{\rm out}$ is the radius of the circumscribing sphere and
\begin{equation}
  r_{\rm mid} = \frac{1}{\sqrt{N\alpha}}
\end{equation}
is the radius of the \emph{mid-sphere}, the sphere around the
barycenter within which any two points always have an inner product
that satisfies the lower bound.

The fundamental ``bit'' in our theory is a system whose state space
$\cP_{\rm bit}$ meets the following condition.  The maximal size of a
set of points, any two of which are maximally distant from each other,
is equal to exactly 2.  In other words, given any hypothesis of
maximal confidence about the system, there exists exactly one other
hypothesis of maximal confidence fully distinct from it.  (And if this
applied only to \emph{some} of those extremal hypotheses, then the
state space would have ``distinguishing marks and scars'' indicating
some other empirical constraint that we are not allowing at this level
of generality.)  For our state space $\cP_{\rm bit}$ to have this
geometrical property and to satisfy the maximality condition given
above, it must be a \emph{ball} within the probability simplex
in~$\bbR^N$.  Moreover, we can say which ball it is, because unless we
invoke further empirical additions to probability theory, we have to
use the distance scale that is already available to us, meaning that
$\cP_{\rm bit}$ is the \emph{inscribed ball of the probability
  simplex.}  That is, to satisfy the condition $d = 2$, we must have
$r_{\rm out} = r_{\rm mid}$, and both of them are in turn fixed to the
value
\begin{equation}
  r_{\rm out} = r_{\rm mid} = r_{\rm in} = \frac{1}{\sqrt{N(N-1)}}.
\end{equation}

We still have to fix the Euclidean dimension $N$ for the fundamental
``bit'' system.  There are multiple ways of doing
so~\cite{Appleby:2017}, each of which suggest different ``foil
theories'' to quantum mechanics.  However, even before fixing $N$, we
know already that as long as the ``nonclassical marginalization'' rule
(\ref{eq:urgleichung}) is indeed nonclassical (that is, $\beta > 0$),
our theory will spurn explanation in terms of intrinsic hidden
variables.  With $N = 4$, we recover the Bloch ball and the theory of
a single qubit, including all the POVMs that can be performed upon it.
This is already enough to prove the impossibility of underlying hidden
variables: While a qubit theory restricted to von Neumann measurements
can be explained classically, fully general POVMs upon a qubit
cannot~\cite{Spekkens:2005}.

In order to reach this conceptual point as quickly as possible, we
have stepped only lightly upon the algebraic manipulations.  For more
detailed treatments, see \cite{Fuchs:2009} and \cite{Appleby:2017}, as
well as~\cite{Fuchs:2016b, Stacey:2019b}.

Having derived the theory of a single qubit, one way forward would be
to compose them.  An argument based on Einstein locality, and using
the postulate of a reference measurement to justify a Gleason-type
noncontextuality condition, gives the tensor-product rule for
composing state spaces~\cite{Fuchs:2002, Barnum:2005}.  This still
leaves open the question of why the joint states for, say, a bipartite
system of two qubits should be the \emph{positive semidefinite}
operators on the tensor-product space.  A voyage into the literature
yields at least one answer to this question~\cite{Barnum:2010,
  DeLaTorre:2012}, but the assumptions involved may feel unsatisfying
--- mathematically over-powered or physically under-motivated.

Another option, and the one we prefer, is to start with
Eq.~(\ref{eq:urgleichung}) and take it as the basic way to express the
nonexistence of intrinsic hidden variables for systems of all
dimensionalities $d$.  In the same manner as before, this yields a
theory defined by a \emph{qplex,} a proper subset of the probability
simplex in~$\bbR^{d^2}$.  Any two points $P$ and $P'$ in a qplex $\cP$
satisfy the inequalities
\begin{equation}
  \frac{1}{d(d+1)} \leq P \cdot P' \leq \frac{2}{d(d+1)}.
\end{equation}
For details on how to go from the very general-looking
Eq.~(\ref{eq:urgleichung}) to these particular inequalities,
see~\cite{Appleby:2017}.  The remaining challenge is then to identify
the qplexes that are isomorphic to quantum state space --- the
\emph{Hilbert qplexes} --- among all the possibilities, and to do so
in the most economical possible way.  Current research focuses on
relaxing the conditions that are known to work, and understanding the
geometrical structures that are necessary to establish the
isomorphism, which turn out to be quite captivating
entities~\cite{Zauner:1999, Renes:2004, Scott:2010, Appleby:2017b,
  Fuchs:2017, DeBrota:2019, DeBrota:2020}.

Assuming that the Law of Total Probability is the way to
``marginalize'' over an unperformed experiment, and the desire for
distinguishable states to be nonoverlapping distributions, are
expressions of faith in hidden variables.  QBism holds that the way to
make progress is to let go of this faith and encourage something more
interesting in its place. This analysis echoes Demopoulos'
criticism of the early Putnam, in which Demopoulos points out how a
worldview can reduce to being a hidden-variable one even if it was not
formulated with a specific ontology --- Bohmian or de Broglian, say
--- explicitly in mind~\cite{Demopoulos:2010}:
\begin{quotation}
  \noindent By claiming to avoid hidden variables, Putnam almost
  certainly meant that he could avoid a theory like Bohm's or de
  Broglie's. For Putnam, such theories are characterized by their
  invocation of special forces to account for the disturbance by
  measurement. Since he nowhere appeals to such forces, Putnam
  believes his account to be free of an appeal to hidden
  variables. The idea that his interpretation of the logical
  connectives assumes an underlying space of truth-value assignments
  --- and that it is therefore a hidden variable interpretation of
  quantum mechanics after all --- appeals to an abstract notion of
  hidden variable theory, one which was only beginning to emerge when
  Putnam wrote his paper, and one with which he was at the time almost
  certainly unfamiliar.
\end{quotation}

\section{Interlude: The PBR Theorem}
In the previous section, we explored the consequences of abandoning
the notion that \emph{distinguishable} means \emph{nonoverlapping.}
This provides a convenient setting to discuss the
Pusey--Barrett--Rudolph (PBR) theorem~\cite{Pusey:2012}. The PBR
theorem is a ``no hidden variables'' argument, or to say it more
precisely, an argument that hidden variables will end up being
redundant. Following the tradition of what nowadays is called the
``ontological models framework''~\cite{Harrigan:2010}, it tries to
express quantum probabilities as being due to ignorance of underlying
ontic states, and then shows that different quantum states must
correspond to nonoverlapping distributions over the ontic state
space. The crucial feature of the ontological models framework is that
probabilities for experiment outcomes are calculated using the Law of
Total Probability:
\begin{equation}
  \tr \rho E = \int d\lambda\, P_\rho(\lambda) P(E|\lambda).
\end{equation}
By adopting the Born Rule in the ``quantum marginalization'' form of
Eq.~(\ref{eq:urgleichung}) as a fundamental postulate, we spurn this
framework from first principles onward. And because QBism lies outside
of the ontological models framework, preferring to find its realism on
a more subtle and stimulating level, the PBR theorem has little to say
about it.  Indeed, the PBR theorem has no \emph{direct} impact on
QBism at all, though it may have the indirect effect of indicating
that half-hearted attempts at interpreting quantum theory in an
informational way do not go far enough.

The PBR theorem relies upon a compatibility criterion between
quantum states that was first codified by Caves, CAF and
Schack~\cite{Caves:2002}, based on an earlier proposal by
Peierls~\cite[p.\ 11]{Peierls:1991}.  Bub has an admirably concise
exposition of the PBR theorem, in an essay where he also raises the
intriguing question of proving analogues of it in nonclassical
theories that are not quantum mechanics. He shows that an analogue of
the PBR theorem can be proved in a nonclassical and nonquantum setting
that he calls ``Bananaworld''~\cite{Bub:2012}.

\section{Demopoulos}

The interpretational work of Demopoulos owes much to Pitowsky.  He
credits Pitowsky for advancing and defending ``the view that the
$\psi$--function represents a state of belief about a system rather
than its physical state''~\cite{Demopoulos:2012}.  Demopoulos regards
the probabilities in quantum physics as being probabilities over
``effects'', which are explicitly \emph{not} the effects that comprise
POVMs~\cite{Demopoulos:2010}. Rather, they ``are to be thought of as
the traces of particle interactions on systems for which we have
`admissible' theoretical descriptions in terms of their dynamical
properties''~\cite{Demopoulos:2012}. It is ultimately difficult to
draw a line between this and, for instance, Pauli's imagery of data
recorded ``by objective registering apparatus, the results of which
are objectively available for anyone's inspection''~\cite{Pauli:1994}.
Therefore, we find Demopoulos' interpretation vulnerable to critiques
in the Wigner's Friend tradition.

That said, there is much in Demopoulos' writing that aligns with
QBism, and in the interests of solidarity and emotional positivity, we
will spend much of this section talking about that.

First, let us get out of the way one of the more significant divisions
between QBism and the position that Demopoulos articulated.  From the
standpoint of the former, the latter tends to slide back to a rather
Bohrian view that quantum phenomena must be treated through classical
intermediary apparatus, with a definition of \emph{classical} that may
have more to do with stable record-keeping than with the details of
Newtonian or Maxwellian dynamics.

Quoting a letter that CAF wrote to Demopoulos on 9 January
2011~\cite{Fuchs:2014}:
\begin{quotation}
  \noindent I didn't see that you strengthened your defense against my
  charge of a kind of dualism.  Maybe I'm still missing something.
  When you write, ``Such systems are epistemically accessible to an
  extent that systems which are characterized only in terms of their
  eternal properties and their effects are not,'' I would think that
  you're admitting that there are (at least) two distinct kinds of
  systems in the world: Those that have a certain type of epistemic
  accessibility and those that do not, and that that epistemic
  accessibility is not characterized by things to do with the
  observer's situation, but rather the with system's
  itself. [\ldots\!] And you don't weaken my accusation any either by
  writing on page 18: ``This kind of conceptual dependence, does not
  preclude the application of quantum mechanics to systems that record
  effects. Although it is largely a matter of convenience which
  systems are, and which are not, taken to record effects, it is not
  wholly a matter of convenience.''  The phrase `not wholly a matter
  of convenience' again points me to an ontic distinction between two
  types of system.  Again, a reason for my saying, ``dualism''.
  Similarly with respect to your sentence, ``This leaves entirely open
  the empirical question of why it is that some systems appear to be
  amenable to descriptions that are expressible in the framework of
  classical mechanics.''  I.e., you don't explain why, but there is an
  empirical distinction between two kinds of system.

I'm not saying there's anything necessarily philosophically wrong with
a dualism; mostly it is that it just doesn't match my taste, and
doesn't feel like the right direction for moving physics forward.  It
adds a bigger burden on the physicist than I'd rather him have: For
now, for each lump of matter, he'll have to --- in some mysterious
way --- come to a conclusion of whether it supports dynamical properties
or not.  How does he do that?  Where you leave us is where Bohr left
us --- as far as I can tell, in just telling us ``it must be so, but I'm
not going to tell you where/how to make that distinction''
\ldots\ i.e., that there must be two kinds of system for us to build
our evidentiary base for the quantumly treatable ones at all.

Imagine my going up to Rainer Blatt and saying, ``Rainer, I just found
this nice rock on the beach.  You think you might be able to use it as
a component in that quantum computer you want to build?''  Asher Peres
would say, ``Of course he can use it as a component; it is just a
matter of money.  With enough money, any old rock can be polished into
a quantum computer.''  But if it ain't so, then it must be a burden on
physics to say when it can and when it cannot be done.  My guess is
that Rainer will never be able to codify and make explicit such a
criterion of distinction; he'll never be able to tell me which tests
he must perform to certify my rock ineligible for quantum computation.
\end{quotation}

When discussing Gleason's work, Demopoulos considers \emph{two-valued
  measures,} maps from the closed subspaces of a Hilbert space to the
pair $\{0, 1\}$.  He notes~\cite{Demopoulos:2012},
\begin{quotation}
  \noindent It is evident that generalized two-valued measures and
  generalized truth-value assignments are formally interchangeable
  with one another, whatever the conceptual differences between
  probability measures and truth-value assignments.
\end{quotation}
QBism stresses the ``conceptual differences'': Its preferred school of
Bayesianism forcefully rejects the idea that a probability-1
assignment indicates pre-existing physical truth.  This leads to
dismissing the EPR criterion of reality.\footnote{Likewise, it implies
  that ``incompatible'' quantum-state assignments do not have to
  indicate different intrinsic physical conditions of a system.  Two
  quantum states $\rho$ and $\rho'$ are incompatible in the Peierlsian
  sense if there exists a measurement for which ascribing $\rho$
  implies the outcome probabilities $(0, 1)$, whereas ascribing
  $\rho'$ implies the outcome probabilities $(1, 0)$.  Shifting one's
  state assignment from $\rho$ to $\rho'$ thus changes a probability-0
  assignment into probability-1, but the agent has no obligation to
  regard this as a shift of a pre-existing physical property, for the
  same reason that they do not take EPR as gospel.} As CAF wrote to
Demopoulos on 29 November 2007~\cite{Fuchs:2014}, regarding the paper
eventually published as~\cite{Demopoulos:2010}:
\begin{quotation}
  \noindent Where I do take issue with what you write is the very
  difficult issue of probability 1.  If I understand you correctly, I
  disagree with the first sentence of the first full paragraph from
  page 22: ``The representation of an elementary particle as a
  function which, when presented with an experimental configuration,
  yields an effect, is interchangeable with its representation as a
  class of propositions only when the effects are predictable with 0-1
  probability.''

For, I would say, not even then.
\end{quotation}
Additional discussion on this point can be found in the
\hyperref[24june2006]{24 June 2006} entry drawn from correspondence
with Bub in Appendix B.

Having established this difference in content and in taste, we can now
proceed to matters of greater agreement. Demopoulos
writes~\cite{Demopoulos:2010},
\begin{quotation}
  \noindent Instead of providing a solution to the long-standing
  issues involving measurement and the paradoxes, the discovery of ``the
  logic of quantum mechanics'' revealed a new and different problem,
  namely, the impossibility of interpreting the probabilities of the
  theory so that every proposition belonging to a particle is
  non-contextually determinately true or false. Rather than refuting
  classical logic, this discovery should be seen as refuting the idea
  that the probabilities are interpretable as the probabilities of
  such propositions. Ironically, the consequences of quantum mechanics
  for logic are almost the precise opposite of what Quine and Putnam
  imagined they might be. If anything, the problem of hidden variables
  upholds the centrality of classical logic in our theorizing about
  the physical world, while allowing that the Boolean algebraic
  structures, which are so closely associated with classical logic, are
  not appropriate for every use of probability in physics.
\end{quotation}
QBism finds much to like here.  Rather than stressing the ``classical
logic'' language, QBism speaks in terms of \emph{internal
  self-consistency}: Within a given scenario fixed by an agent's
choice of action, the agent's mesh of beliefs should hold together,
i.e., they should be Dutch-book coherent.  Moreover, the notion
of Dutch-book coherence is, after a fashion, built upon ``classical
logic''.  That is, one way of expressing the basic precept of it is
that an agent should not assign different probabilities to events
whose descriptions can be translated to one another by Boolean
manipulations. The Law of Total Probability itself follows from
Dutch-book coherence, but it is ``not appropriate for every use of
probability in physics''.

QBism has been accused of being ``instrumentalism'' with tiresome
regularity.  In this regard, it is helpful to recall Demopoulos'
response to that charge being leveled at his own interpretation.
Replying to the accusation that his view would make quantum physics
``a mere heuristic for prediction'', Demopoulos
noted~\cite{Demopoulos:2010},
\begin{quotation}
\noindent What is at issue is the interpretation of the quantum
algorithm for assigning probabilities and for reasoning from the
position of uncertainty which such probabilities necessarily
signify. This leaves a considerable degree of realism intact. First,
there is nothing in the view that denies the existence of elementary
particles [\ldots\!]. Secondly, the view allows that there is a
plethora of properties which elementary particles are
unproblematically represented as having; these include all of their
eternal or non-dynamical properties. Both they and the particles
themselves are real objects of theoretical investigation.
\end{quotation}
QBism reserves the right to quarrel with the letter of this creed, but
it is in accord with the spirit of it.\footnote{For example, a working
  field theorist might devalue the word \emph{particle,} or at least
  drop many connotations of it.  But her motivation to do so would be
  based on the Unruh effect or on spacetime curvature invalidating a
  choice of positive-frequency subspace~\cite{Wald:1994}.  These
  reasons would have little to do with the ``interpretation of quantum
  mechanics'' as that topic has socially been defined.  For more on
  QBism and QFT, see~\cite{DeBrota:2018}.} The technical work outlined
in \S\ref{sec:tech} above is a fundamentally realist investigation of
exactly the type that Demopoulos indicates.

\section{Conclusions}
In this paper, we have availed ourselves of Pitowsky's program and its
extensions, particularly the work of William Demopoulos, to better
explicate QBism's aims and goals.

The Pitowskyan interpretation of quantum mechanics is, chiefly, the idea
that the formal structure of quantum theory should be seen as a
Bayesian probability theory, adapted to the empirical situation that
Nature's events conspire to conform to a non-Boolean algebra.
QBism also gives a Bayesian reading of quantum probabilities, but it
instead regards them as the personal degrees of belief a
savvy agent might hold for personal experiences
arising from her actions on the external world.  Thus QBism has two
levels of personalism, whereas the Pitowskyan view has only one.

As we have seen, the differences go further and get into the technical
meat of quantum information.  The Born Rule is crucially important for
the technical developments of both views, but in the Pitowskyan
tradition it is a theorem, derived per Gleason from the underlying
assumption of a non-Boolean algebra.  In contrast, QBism strives to
place the Born Rule in a pre-algebraic setting, so that it itself may
shine forth as the primary empirical statement of the theory.  The
QBist approach to reconstructing quantum theory hopes that, in the
right language, the Born Rule is quantum theory's most fundamental
postulate, with the Hilbert space formalism arising as a consequence.
There remains a certain \emph{\`a la carte} aspect to the
reconstruction of quantum theory from the Born Rule; the aspiring
reconstructor has at some junctures her choice of secondary postulates
to invoke, though they mostly have the character of asking that the
resulting structures be as mathematically unremarkable as possible so
that all the \emph{physics} can be loaded into the Born Rule.  We have
stepped through this in the special case of a single qubit and the
POVMs thereupon.

The distinctions between QBism and the Bub--Demopoulos--Pitowsky
school manfiest in shifts of technical emphasis.  We have seen, for
example, how the QBist reconstruction program plays down Gleason's
theorem and the quantum-logic tradition. On a deeper and more
conceptual level, the two diverge in their theories of truth. There
is, at least implicitly, a sentiment for a correspondence theory of
truth in how Bub, Pitowsky and Demopoulos approach the quantum.  By
contrast, QBism invokes a Jamesian, pragmatist attitude. This is a
move that is unfamiliar to many in the present-day
philosophy-of-physics circles, as evidenced by Timpson's needing to
define it for the professionals who read
\booktitle{Stud.\ Hist.\ Phil.\ Mod.\ Phys.}~\cite{Timpson:2008} with
a footnote:
\begin{quotation}
  \noindent Pragmatism is the position traditionally associated with
  the nineteenth and early twentieth century American philosophers
  P[ei]rce, James and Dewey; its defining characteristic being the
  rejection of \emph{correspondence} notions of truth in which truths
  are supposed to mirror an independently existing reality after which
  we happen to seek, in favour of the thought that truth may not be
  separated from the process of enquiry itself. The caricature slogan
  for the pragmatist's replacement notion (definition) of truth is
  that `Truth is what works!'~in the business of the sincere and open
  investigation of nature.
\end{quotation}
(The authors admit a fondness for James, not just on account of his
ideas but also because he writes to be understood --- indeed, to make
the life of the mind a thing that is \emph{felt.})

This turn to pragmatism has concrete expression in the QBist rejection
of the EPR criterion and sidelining of the ontological models
framework.  To a QBist, trying to recast all quantum probabilities as
ordinary, neoclassical ignorance is to miss the point entirely.  One
way or another, such efforts end as baroque restatements of their
starting point.  Much more intriguing are nontrivial reproductions of
\emph{portions} of quantum theory within such a framework, since these
illustrate how some ``quantum'' behaviors are only \emph{weakly
  nonclassical.} Among these are the no-cloning theorem and the fact
that there is generally no good Boolean ``and'' operation for
experiments --- for spin-$\hat{x}$ and spin-$\hat{z}$ measurements
upon a bit system, to pick an easy example.  In turn, this
demonstration indicates that quantum logic put its emphasis in the
wrong place. This is no slight to its pioneers, only a retrospective
judgment based on discoveries after their time.  (As scientists, we
belong to the last profession of romantics~\cite{Gleick:1992}, the
last to believe in geniuses --- and John von Neumann was certainly one
of those.)

Like we are with many subjects, we are quite willing to raid the
quantum-logic bookshelf to find inspiration for technical
conjectures~\cite{Stacey:2019d}, but non-Booleanity is not where we
are seeking the fundamental lesson of quantum mechanics.  For us it is
a consequence, not a presumption.  More broadly, writing the Born Rule
as Eq.\ (\ref{eq:urgleichung}) is part of the attempt to understand
the mathematical foundations of quantum mechanics. In particular, it
helps to make the distinction between its physical content and
mathematical artifact clear.

\section{Acknowledgments}
This research was supported in part by the John Templeton Foundation.  The opinions expressed in this publication are those of the authors and do not necessarily reflect the views of the John Templeton Foundation.  CAF was further supported in part by the John E. Fetzer Memorial Trust, and grants FQXi-RFP-1612 and FQXi-RFP-1811B of the Foundational Questions Institute and Fetzer Franklin Fund, a donor advised fund of Silicon Valley Community Foundation.

\appendix
\section{Even Orthogonal Quantum States are Overlapping Probability
  Distributions}
\label{app:ortho}
We begin by generalizing Eq.~(\ref{eq:qubit-sic}) to an arbitrary
four-outcome reference measurement $\{H_i\}$ for a qubit.  This is the
minimum number of outcomes necessary for a qubit reference
measurement; any fewer, and the POVM elements could not span the state
space~\cite{DeBrota:2019b}.  Given any qubit state $\rho$, at most one
element of the set $\{H_i\}$ can be orthogonal to it, and so at most
one entry in the vector $P(H)$ can equal zero.  Consequently, any two
qubit states $\rho$ and $\rho'$ will have probabilistic
representations with overlapping support.

To generalize further and encompass systems of arbitrary dimension, as
well as reference measurements that are not necessarily minimal, first
we show that we can focus our attention on pure states.  We can write
any density matrix as a convex combination of rank-1 projectors, say
in the matrix's eigenbasis, so if we have two quantum states
\begin{equation}
  \rho := \sum_i c_i \ketbra{\psi_i}{\psi_i},
  \ \rho' := \sum_j c'_j \ketbra{\psi'_j}{\psi'_j},
\end{equation}
then their Hilbert--Schmidt inner product is just a weighted average:
\begin{equation}
  \tr \rho\rho' = \sum_{ij} c_i c'_j \left|\braket{\psi_i}{\psi'_j}\right|^2.
\end{equation}
Let us say that the probabilistic representations of these basis
vectors are
\begin{equation}
  s_i(k) := \tr H_k \ketbra{\psi_i}{\psi_i},
  \ s'_j(k) := \tr H_k \ketbra{\psi'_j}{\psi'_j}.
\end{equation}
Then, the inner product of $\rho$ and $\rho'$ is a weighted average of
the ``$B$-inner products'' of these probability vectors:
\begin{equation}
  \tr \rho\rho' = \sum_{ij} c_i c'_j\, s_i^{\rm T} B s'_j,
\end{equation}
where the matrix $B$ is the inverse of the Gramian of the reference
measurement:
\begin{equation}
  [B^{-1}]_{mn} = \tr H_m H_n.
\end{equation}
The $B$-inner product of two valid probability vectors is bounded
below by zero.  What does this imply for the ordinary Euclidean inner
product, i.e., the dot product?  In particular, can the dot product
ever be zero itself?  To understand this, it suffices to consider two
pure states, since per the above discussion, mixing cannot lower their
inner products.  Suppose that $\ket{\psi}$ and $\ket{\psi'}$ are two
different pure states in dimension $d$.  They define a subspace of
the full space of Hermitian operators on $\bbC^d$.  Let $H_k^{\rm P}$
denote the images of the reference-measurement effects projected into
this subspace.  Any quantum state living within this subspace will
have the same inner products with the $H_k^{\rm P}$ as it did with the
original $H_k$.  Thus,
\begin{equation}
  s(k) := \bra{\psi} H_k \ket{\psi} = \bra{\psi} H_k^{\rm P}
  \ket{\psi},
\end{equation}
and likewise for $s'$, the probabilistic representation of
$\ket{\psi'}$.

Imagine that $s \cdot s'$ were zero.  Because each entry in either
vector is nonnegative, this can only happen when $s$ and $s'$ have
completely disjoint supports.  In other words, some of the $H_k^{\rm
  P}$ will be orthogonal to $\ketbra{\psi}{\psi}$, and some will be
orthogonal to $\ketbra{\psi'}{\psi'}$. And for each value of $k$, at
least one of these options will obtain.

In the qubit-sized subspace defined by $\ket{\psi}$ and $\ket{\psi'}$,
there is a unique state orthogonal to $\ket{\psi}$ and a unique state
orthogonal to $\ket{\psi'}$.  Call these $\ket{\psi_\perp}$ and
$\ket{\psi'_\perp}$; they are antipodal to $\ket{\psi}$ and
$\ket{\psi'}$ respectively on the Bloch sphere for this subspace.
Each of the $H_k^{\rm P}$ must be proportional either to
$\ket{\psi_\perp}$ or to $\ket{\psi'_\perp}$.  But this means that the
$H_k^{\rm P}$ cannot span the subset of state space given by
$\ket{\psi}$ and $\ket{\psi'}$ --- the reference measurement cannot be
``informationally complete'' on this subspace, for any such
measurement must have at least \emph{four} distinct outcomes to cover
a qubit.  Consequently, the proposal that $s \cdot s' = 0$ contradicts
the assumption that we had a working reference measurement in the
first place.  And so, for any reference measurement $H_k$ whatsoever,
the dot product of valid probability vectors will always be greater
than zero, even when the corresponding quantum states are orthogonal.

\section{Additional Excerpts of Correspondence}
\phantomsection
\label{5may2001}
\noindent {\bf Itamar Pitowsky to CAF, 5 May 2001:}
\begin{quotation}
\noindent Of course, you can print [our correspondence]. The whole
thing is a wonderful idea.  Historians of science often complain that
published articles never tell even half the story of science, because
they don't let you see the false starts, misleading intuitions,
errors, or even how a sound idea is baked.  Correspondence makes the
story come alive, but it usually takes almost a century to publish,
and it's done only in cases of the likes of Einstein. \ldots\ Your
collection arrives on the scene almost in real time, very nice.
\end{quotation}
\begin{comment}
[?? Delete?] CAF to Jeff Bub, 18 July 2001:
\begin{quotation}
\noindent I know I suggested I would write a longer letter soon, but
I'm going to wimp out of it again for now.  It would concern the main
point of distinction I see between us (and also between myself and
Pitowsky).  Namely, A) that I view a large part of quantum mechanics
as merely classical probability theory (which on my view may be an a
priori ``law of thought'') PLUS an extra assumption narrowing down the
characteristics of the phenomena to which we happen to be applying it
to at the moment, while B) you are more tempted to view quantum
mechanics as a {\it generalization\/} of classical probability theory
(and with it information theory).
\end{quotation}
\end{comment}
\phantomsection
\label{26june2002}
{\bf CAF to Itamar Pitowsky, 26 June 2002:}
\begin{quotation}
\noindent I too used to think that the [partial Boolean algebra]
approach was the way to go if one wanted to build up a theory along
quantum logical lines. But now, I'm not so convinced of it. That is
because I am starting to think that quantum mechanics is more
analogous to the epistemological theory Richard Jeffrey calls
``radical probabilism'' than anything else. From that view, there are
``probabilities all the way down'' with one never getting hold of the
truth values of {\it any\/} propositions. R\"udiger Schack and I just
discovered a wealth of material on Jeffrey's webpage
\myurl[http://www.princeton.edu/~bayesway]{http://www.princeton.edu/$\sim$bayesway/}.

In any case, I think what this leads to is that we ought to be
focusing much more on characterizing quantum mechanics solely in terms
of the ``logic'' of POVMs than anything else --- these being the
structures analogous to what crops up in Jeffrey's ``probability
kinematics.''  Thus, if one is looking to characterize PBAs, the best
task might be to focus on the kinds of PBAs that POVMs generate,
rather than the ones of Kochen and Specker based solely on standard
measurements.  [\ldots\!] Beyond that, I am now of the mind that all
one really ever needs for understanding quantum mechanics is a {\it
  single\/} Boolean algebra that is kept safely in the background
(solely) for reference.  The rest of the theory (and indeed all
real-world measurement) is about probability kinematics with respect
to that algebra.
\end{quotation}

\phantomsection
\label{21april2004}
{\bf CAF to Bill Demopoulos, 21 April 2004:}
\begin{quotation}
  \noindent As you argued in your earlier letter to me (one from last
  year sometime), our views --- or maybe just our languages --- may not be
  so incompatible as one might think.  However, I am left with the
  feeling that this is only a contingent feature of the particular
  stages of the game we happen to be at, at the moment.  In
  particular, from my own view, I think it is quite important that we
  strive to stop thinking of quantum states as states of knowledge
  about the \emph{truth value} of this or that proposition (even if truth
  value is not invariant with respect to `experimental
  arrangement' --- the idea you are toying with).  My feeling is that
  the imagery of measurement outcomes mapping to truth values (in this
  context anyway) will only cloud our vision for how to take the next
  big step.

What is the next big step?  I think it is a deeper understanding of
how --- very literally --- the world ``is in the making'' (to use a
Jamesian phrase).  To try to make that idea at least graspable (if not
either clear or consistent yet), and to try to show you quantum
theory's role in all this, let me attach four letters I've written
recently. They're contained within the attached files.\footnote{In the large
  samizdat~\cite{Fuchs:2014}, see the 24 June 2002 note ``The World is
  Under Construction'' to Wiseman; the 27 June 2002 note ``Probabilism
  All the Way Up'' to Wiseman; the 12 August 2003 note ``Me, Me, Me''
  to Mermin and Schack; and the 18 August 2003 note ``The Big IF'' to
  Sudbery and Barnum.}  I think they are my best
statements to date of what I am shooting for; and I think that goal
fundamentally conflicts with the idea of ``measurement'' propositions
having truth values in the conventional sense.

That is not to say, however, that I am yet ready to give up on the
idea of physical systems having autonomous properties.  The question
is, what can still be pinned down as a property in the conventional
sense?
\end{quotation}

\begin{comment}
[?? Delete?] CAF to Bill Demopoulos, 6 May 2004:
\begin{quotation}
  \noindent What I would like to go into further eventually is the
  similarity and differences between 1) your point of view of the
  underlying events in QM as being preexistent (but maybe unique)
  realities, and 2) my point of view that they are best supposed as
  (unique) creations.  I also get the feeling that your view collides
  with another one of my favorite doctrines: Namely, that we should
  take POVMs as just as basic as standard quantum measurements
  (because they so prettily match Bayes' rule).  But I'll have to
  think about that.
\end{quotation}

[?? Delete?] CAF to Bill Demopoulos, 2 January 2006:
\begin{quotation}
  \noindent The transformation that quantum mechanics speaks about,
the transformation from a `superposition' to `aliveness' or
`deadness', is a transformation {\it within the agent}, and that
transformation cannot take place without some interaction with the
external physical system labeled by the word `cat'.  What happens to
`cat' itself (described in a way that makes no reference to the
agent)?  On that, I think quantum mechanics is silent.  With a
mantra:  Quantum mechanics is a theory for ascribing (and
intertwining) personal probabilities for the personal consequences of
one's personal interactions with the external world.
\end{quotation}
\end{comment}

\phantomsection
\label{17february2006}
{\bf CAF to Bill Demopoulos, 17 February 2006:}
\begin{quotation}
  \noindent Let me try to consider a situation and 1) try to imagine
  what you would say of it (but probably in my idiosyncratic
  language), followed by 2) what I think I would say of it \ldots\ and
  then see if there is a substantial distinction.

Start with a finite dimensional Hilbert space, say of dimension 3,
and imagine it indicative of some real physical system within an
observer's concern.  From that Hilbert space, let us form all
possible sets of three mutually orthogonal one-dimensional projection
operators.  That is, let us consider all possible sets of the form
$\{P_1, P_2, P_3\}$.

What is it that you would say of those sets?  If I understand you
correctly, it is this.  Each such set $\{P_1, P_2, P_3\}$ corresponds
to a set of mutually exclusive properties that the system can
possess.  At any given time, one of those projectors will have a
truth value 1 and the other two will have values 0.  Now consider a
potentially different such set $\{Q_1, Q_2, Q_3\}$; again, at any
given time, one of those projectors will have a truth value 1 and the
other two will values 0.  What is interesting in your conception, if
I understand it, is that even if two elements happen to be identified
between those two sets --- for instance, if $P_1=Q_3$ --- there is {\it
no requirement\/} that $P_1$ and $Q_3$ need have the same truth
value; $P_1$ might have the truth value 0, whereas $Q_3$ might have
the truth value 1.  Another way to say this is that the truth-value
assignments depend upon the whole set and not simply the individual
projection operators.  For you, all the identification $P_1=Q_3$
amounts to is that the {\it probability\/} for the truth value of
$P_1$ within the set $\{P_1, P_2, P_3\}$ is the {\it same\/} as the
{\it probability\/} for the truth value of $Q_3$ within the set
$\{Q_1, Q_2, Q_3\}$. (If you were a Bayesian about
probabilities --- though I don't think you are --- you would say, ``Well
$P_1$ has whatever truth value it does, and $Q_3$ has whatever truth
value it does (each within their appropriate set of mutually
exclusive triples), but my degree of belief about the truth value of
$P_1$ is the same as my degree of belief about the truth value of
$Q_3$.  That is the rule I am going to live by.'')  Then it follows
from Gleason's theorem that there exist no probability assignments
for the complete (i.e., continuously infinite) set of triples that are
not of the quantum mechanical form. In particular, one can never
sharpen one's knowledge to a delta function assignment for {\it
each\/} triple. This is how you cash out the idea of an `incompletely
knowable domain.'

That is a novel idea, and if I understand it correctly, I like it.

However, now let me contrast my characterization of you with what I
think has been my working conception.  I prefer not to think of the
triples $\{P_1, P_2, P_3\}$ as sets of mutually exclusive {\it
properties\/} inherent within the system all by itself, but rather
{\it actions\/} that can be taken upon the system by an external {\it
agent}. Each {\it set\/} of such projectors corresponds to a distinct
action; what the individual elements within each set represent are
the (generally unpredictable) {\it consequences\/} of that action.
What are the consequences in operational terms?  Distinct sensations
within the agent.  The reason I insist on calling them consequences,
rather than ``sensations'' full stop, is because I want to make it
clear that the domain of what we are talking about is sensations that
come about through the action of an agent {\it upon\/} the external
world.

The essential idea of [what was to become QBism]
is that no element of a set $\{P_1, P_2, P_3\}$ has a truth value
before the action of the agent.  Rather the truth value --- if you want
to call it that (maybe it is not the best terminology) --- is generated
(or given birth to) in the process.  At that point, one of the $P_i$
stands in autonomous existence (within the agent), whereas the other
two fall.

I hope I have characterized both of us accurately!

Here is the question that has been troubling me.  Is there any real
distinction (one that makes an pragmatic difference) between our
views?  You say the truth value is there and revealed by the
measurement, and I say it's made by the measurement and wasn't there
beforehand.  So what?

If there is a pragmatic distinction, Steven van Enk and I through
discussions this week have come to believe that it may show up most
clearly in how you and I would treat counterfactuals with regard to
measurement.  Let us take a situation where an agent ascribes a
quantum state $\rho$ to the system; contemplating the measurement
$\{P_1, P_2, P_3\}$, we know that he will ascribe probabilities
according to the Born rule $\tr(\rho P_i)$ for the various
outcomes. Suppose he now performs that measurement and actually gets
value $P_2$.

What does getting that outcome teach him about the quantum system?  I
think you would say it reveals which of the three mutually exclusive
properties the system actually had.  On the other hand, I would say
it teaches him nothing about the system per se; the outcome $P_2$ is
just the consequence of his action.  What is the implication of this
on counterfactuals?  Here's at least one.

Suppose after you get your outcome, you contemplate magically having
performed a distinct measurement $\{Q_1, Q_2, Q_3\}$ instead.  I
think you're careful to point out in your paper that the knowledge of
$P_2$ carries no implication for what you would have found with this
other imaginary measurement.  But what happens if you conceptually
transform this measurement $\{Q_1, Q_2, Q_3\}$ to one closer and
closer to the original, i.e., to $\{P_1, P_2, P_3\}$?  In the {\it
limit\/} when the two are identical again, I think you would say that
knowledge of the outcome $P_2$ in the original case implies that
$P_2$ will also be the outcome in the limiting counterfactual case.
But what would I say?  From my conception, there is no reason at all
to believe that the limiting counterfactual case will give rise to
the same outcome $P_2$.  The best one can do, either in the original
case or the counterfactual case, is to say that an outcome $i$ will
arise with probability $\tr(\rho P_i)$.  In fact, a
counterfactual analysis with this kind of result may be the very
meaning of the idea that quantum measurements are generative of their
outcomes.
\end{quotation}

\begin{comment}
[?? Delete?] CAF to Bill Demopoulos, 18 February 2007:
\begin{quotation}
  \noindent 1) Almost by definition, what you are proposing is a
  contextual hidden variable theory.  But what is its status with
  regard to locality?  At different times (while driving, taking a
  shower, etc.), I've been able to convince myself that a constraint
  of {\it locality\/} can be placed upon the truth values, but then I
  get confused.  What can you say on the matter?

2)  Take two triads of one dimensional projectors, $\{P_1, P_2,
P_3\}$ and $\{Q_1, Q_2, Q_3\}$, as in my last note to you.  And as
before, suppose $P_1=Q_3$.  However this time, let us be careful to
assume that $P_1$ and $Q_3$ differ in truth value in their respective
sets.  What happens now when we consider nonelementary propositions
of the form $\{P_1, \neg P_1\}$ and $\{\neg Q_3, Q_3\}$ where by
$\neg P_1$ I mean the orthocomplement of $P_1$, etc.  Presumably you
still want to view these sets as representative of mutually exclusive
properties inherent within the quantum system.  However, by
construction the sets are identical: $\{P_1, \neg P_1\} = \{\neg Q_3,
Q_3\}$. How does one decide on a truth value assignment here, given
the previous truth value assignments for $\{P_1, P_2, P_3\}$ and
$\{Q_1, Q_2, Q_3\}$?
\end{quotation}
\end{comment}

\phantomsection
\label{24june2006}
{\bf CAF to Jeffrey Bub, 24 June 2006:}

\vspace{-24pt}
\begin{quotation}
\bjb
The transition from classical to quantum mechanics involves replacing
the representation of properties as a Boolean lattice, i.e., as the
subsets of a set, with the representation of properties as a certain
sort of non-Boolean lattice.
\ejb

\vspace{12pt}

The?  I would rather say {\it one possible way of looking at\/} the
transition from classical to quantum mechanics involves blah, blah,
blah. And, you partially recover from this a few paragraphs later
where you write:

\bjb \label{Bubism5}
Of course, other ways of associating propositions with features of a
Hilbert space are possible, and other ways of assigning truth values,
including multi-valued truth value assignments and contextual truth
value assignments. Ultimately, the issue here concerns what we take
as the salient structural change involved in the transition from
classical to quantum mechanics, and this depends on identifying
quantum propositions that take the same probabilities for all quantum
states.
\ejb
But let me hang on this point for a moment despite your partial
recovery. For when you say things like, ``Fuchs misses the essential
point,'' you should realize that that judgement (at most) comes from
within a context very different from the one I am working in.

I would, for instance, never say ``the representation of properties
in quantum mechanics involves as a certain sort of non-Boolean
lattice.''  That is just not the context I'm working in.  Similarly,
I would not say, as you say in the next section, ``Somehow, a
measurement process enables an indeterminate property, that is
neither instantiated nor not instantiated by a system in a given
quantum state, to either instantiate itself or not with a certain
probability.'' --- i.e., I would not say that a measurement process
instantiates any {\it property\/} at all for a quantum system.

Instead, the setting for our quantum Bayesian program (i.e., the
particular one of Caves, Schack, and me), is one where all the  {\it
properties\/} \underline{intrinsic} to a quantum system are timeless
and have no dynamical character whatsoever --- moreover, those
properties have nothing to do with particular quantum state
assignments or particular quantum measurement outcomes. In that way,
the idea of a non-Boolean lattice simply doesn't apply to them.

John Sipe recently made a nice write-up of our view for his book
that, I think, brings this one difference between you and me into
pretty stark relief.  Maybe it's worthwhile to quote it at length, as
it may lay the groundwork for a good bit of our later discussion:

\bq
This interpretation shares some features with operationalism. {\it
Measurements}, for example, are understood in a manner close to that
adopted by an operationalist. They are characterized by POVMs, and
those abstract elements are associated with tasks in the laboratory
undertaken with gadgets that are part of the primitives of the
theory. The result of any such measurement is simply one of a
possible number of outcomes, and there is no talk of these
measurements ``revealing'' the value of any variable, in the sense
that an arbitrarily precise position measurement in classical
mechanics is often described as revealing the position of a particle.
Yet, compared to the operationalist's quiet, unassuming terminology
of ``tasks'' and ``outcomes,'' advocates of this interpretation adopt
a more active manner of speaking, referring to ``actions'' (or even
``interventions'') undertaken by an agent, and the ``consequences''
that those actions elicit.

This indicates a role for the observer (or agent) in this
interpretation that is more significant than the role played by such
a person in operational quantum mechanics. The significance of that
role becomes clear when we consider the reference of density
operators in this interpretation. Density operators do not refer to
sets of tasks that define {\it preparations}, as they do in
operational quantum mechanics. Rather, a density operator is taken to
encode the beliefs of an agent concerning the probabilities of
different consequences of possible future actions. While these
beliefs may be {\it informed\/} by knowledge of the tasks involved in
setting up the particular gadgetry associated with a preparation,
they are not {\it determined\/} by it. Hence there is not a unique,
``correct'' density operator necessarily associated with each
preparation procedure, as there is in operational quantum mechanics.
In the present view two different researchers, one more skilled in
quantum mechanics than the other, could adopt different density
operators after being identically informed of the details of a
particular preparation procedure. One density operator might be more
successful than the other in predicting the possible consequences of
future actions, but each would be the correct density operator {\it
for that agent\/} insofar as it correctly encoded that agent's
beliefs.

Thus, while the abstract elements in the theory associated with
measurements are identified with tasks in the laboratory, as in
operationalism, the abstract elements in the theory associated with
preparations are identified with beliefs of the agent, signaling a
kind of empiricist perspective.

So in contrast to operational quantum mechanics, where density
operators are necessarily updated following a measurement  ---  since
the combination of the previous preparation and the measurement
constitutes a new preparation, and an operationalist associates the
new density operator with {\it that} --- in this view there is no
necessary updating of a density operator in the light of measurement
outcomes, since there is no {\it necessary\/} connection between the
consequences of an agent's action (more prosaically, ``measurement
outcomes'') and his or her beliefs. After all, foolish researchers,
like foolish men and women more generally, could choose not to modify
their beliefs concerning the consequences of future actions despite
their knowledge of the consequences of recent ones. And note that
even wise researchers will not update their beliefs concerning future
actions until they {\it know\/} the consequences of recent ones;
hence a wise researcher's ``personal density operator'' (the only
kind of density operator there is in this view!)\ will not change
until that researcher is actually aware of a measurement outcome.

Other abstract elements in the theory, such as the dimension of the
Hilbert space, and the dimensions of various factor spaces, are
actually associated with instances of attributes of physical objects.
Hence with respect to the reference of {\it these\/} abstract
elements this interpretation is realist. The manner in which this
works can best be seen by first reviewing the role measurement
outcomes play in revealing aspects of the universe in realist
classical mechanics, and then comparing that with the role such
outcomes play in this interpretation of quantum mechanics.

An arbitrarily precise position measurement of a bead moving along a
wire, in realist classical mechanics, reveals the position of the
particle, the instance (say, $x = 10$\,cm) of a particular attribute
(bead position) of a physical object (bead) that actually exists in
Nature. In contrast, a usual Stern--Gerlach device oriented along the
z direction {\it does not}, in this interpretation of quantum
mechanics, reveal the z-component of angular momentum, or for that
matter anything else. The particular outcome of one experimental run
is simply a consequence of performing the experiment. Nonetheless,
repeated experimentation {\it does\/} reveal that the electron
associated with the atom passing through the device should be taken
as a spin-1/2 particle. Here the attribute under consideration is
taken to be {\it internal angular momentum}, and the instance -- the
irreducible representation appropriate to the particle of interest --
{\it spin-1/2}. The role of an ``instance of an attribute'' in this
interpretation is not to specify one of a number of possible
expressions of existence, as it is in realist classical mechanics,
but rather to specify one class of possible beliefs -- the one that
the theory recommends -- about the consequences of future
interventions of a particular type.

Note that, at least within nonrelativistic physics, the instances of
the attributes in this interpretation are fixed. A spin-1/2 particle
remains a spin-1/2 particle. Thus there are no dynamical variables in
this theory, only nondynamical variables analogous to the mass of a
particle in nonrelativistic classical mechanics. The point of physics
is to identify these nondynamical variables. Repeated interventions
by experimentalists, and the careful noting of the range of
consequences that those interventions elicit, is how these fixed
instances are discovered.

In this interpretation of quantum mechanics, with its mix of
operationalist, empiricist, and realist identification of abstract
elements in the theory, these fixed instances specify the [[agent
independent features]] of the ``quantum world,'' and it is the
business of physics to figure them out. This is done by
experimentation, and the theoretical linking of basis vectors in the
appropriate Hilbert space with various measurements, providing an
``anchor'' for those basis kets to our experience, the consequences
of our actions. Particularly significant is the Hamiltonian operator
and its basis kets [[in Caves' particular version of all this]]. As
time evolves during what is colloquially described as ``unitary
evolution,'' we have the option to modify our beliefs or to modify
the anchors of those beliefs; the first strategy corresponds to the
usual Schr\"odinger picture, the second to the Heisenberg picture.

Regardless of the strategy, the [[properties intrinsic to the]]
quantum world of this interpretation [[are]] a fixed, static thing.
[[This aspect of the quantum world]] is a frozen, changeless place.
Dynamics refers not to the quantum world, but only to our actions,
our experiences, and our beliefs as agents. Or, more poetically (\`a la
Chris), life does not arise from our interventions; it is our
interventions.
\eq

John\index{Sipe, John E.} doesn't represent us correctly in every detail of this
presentation --- for the purpose at hand, it only seemed essential to
modify him in a few instances, which I have have marked with double
brackets [[$\bf \cdot$]] --- but I would say he is roughly on track,
and he certainly gets it that we are not concerned with the usual way
of ascribing properties to quantum systems via the values of
measurement outcomes or probability-1 predictions (i.e., the
eigenvector-eigenvalue link).

Which brings me back again to your paper:
\bjb \label{Bubism6}
For a quantum state, the properties represented by Hilbert space
subspaces are not partitioned into two such mutually exclusive and
collectively exhaustive sets: some propositions are assigned no truth
value. Only propositions represented by subspaces that contain the
state are assigned the value `true,' and only propositions
represented by subspaces orthogonal to the state are assigned the
value `false.' This means that propositions represented by subspaces
that are at some non-zero or non-orthogonal angle to the ray
representing the quantum state are not assigned any truth value in
the state, and the corresponding properties must be regarded as
indeterminate or indefinite: according to the theory, there can be no
fact of the matter about whether these properties are instantiated or
not.
\ejb
You see, my way of looking at things wouldn't even allow me to say
what you say here.  It is just a very different world that I am
working in.

To try to make this point, let me quote a couple of emails I wrote to
Bas van Fraassen a few months ago.  It started with my saying this:
\bq
The way I view quantum measurement now is this.  When one performs a
``measurement'' on a system, all one is really doing is taking an
\emph{action} on that system.  From this view, time evolutions or unitary
operations etc., are not actions that one can take on a system; only
``measurements'' are.  Thus the word measurement is really a
misnomer --- it is only an action.  In contradistinction to the old
idea that a measurement is a query of nature, or a way of gathering
information or knowledge about nature, from this view it is just an
action on something external --- it is a kick of sorts.  The
``measurement device'' should be thought of as being like a
prosthetic hand for the agent --- it is merely an extension of him; in
this context, it should not be thought of as an independent entity
beyond the agent.  What quantum theory tells us is that the formal
structure of all our possible actions (perhaps via the help of these
prosthetic hands) is captured by the idea of a
Positive-Operator-Valued Measure (or POVM, or so-called ``generalized
measurement'').  We take our actions upon a system, and in return,
the system gives rise to a reaction --- in older terms, that is the
``measurement outcome'' --- but the reaction is in the agent himself.
The role of the quantum system is thus more like that of the
philosopher's stone; it is the catalyst that brings about a
transformation (or transmutation) of the agent.

Reciprocally, there [[may]] be a transmutation of the system external
to the agent.  But the great trouble in quantum interpretation --- I
now think --- is that we have been too inclined to jump the gun all
these years:  We have been misidentifying where the transmutation
indicated by quantum mechanics (i.e., the one which quantum theory
actually talks about, the ``measurement outcome'') takes place.  It
[[may]] be the case that there are also transmutations in the
external world (transmutations in the system) in each quantum
``measurement'', BUT that is not what quantum theory is about.
[[Quantum mechanics]] is only a hint of that more interesting
transmutation.   [[Instead, the main part of quantum mechanics is
about how]] the agent and the system [[together bring about]] a
little act of creation that ultimately has an autonomy of its
own.
\eq
which led to the following dialogue:

\vspace{-18pt}
\bq
\bvf
Writers on the subject have emphasized that the main form of
measurement in quantum mechanics has as result the value of the
observable at the end of the measurement -- and that this observable
may not even have had a definite value, let alone the same one,
before.
\evf

\vspace{8pt}

Your phrase ``\emph{may not} even have a definite value'' floated to my
attention.  I guess this floated to my attention because I had
recently read the following in one of the Brukner/Zeilinger papers,
\bq\noindent
     Only in the exceptional case of the qubit in an eigenstate of
     the measurement apparatus the bit value observed reveals a
     property already carried by the qubit.  Yet in general the value
     obtained by the measurement has an element of irreducible
     randomness and therefore cannot be assumed to reveal the bit
     value or even a hidden property of the system existing before
     the measurement is performed.
\eq
I wondered if your ``may not'' referred to effectively the same thing
as their disclaimer at the beginning of this quote.  Maybe it
doesn't. Anyway, the Brukner/Zeilinger disclaimer is a point that
Caves, Schack, and I now definitely reject:  From our view all
measurements are generative of a \emph{non}-preexisting property regardless
of the quantum state.  I.e., measurements never reveal ``a property
already carried by the qubit.''  For this, of course, we have to
adopt a Richard Jeffrey-like analysis of the notion of
``certainty'' --- i.e., that it too, like any probability assignment,
is a state of mind --- or one along (my reading of)
Wittgenstein's --- i.e., that ``certainty is a tone of voice'' --- to
make it all make sense, but so be it.
\eq
and

\vspace{-18pt}
\bq
\bvf
Suppose that an observer assigns eigenstate $|a\rangle$ of $A$ to a
system on the basis of a measurement, then predicts with certainty
that an immediate further measurement of $A$ will yield value $a$,
and then makes that second measurement and finds $a$.  Don't you even
want to say that the second measurement just showed to this observer,
as was expected, the value that $A$ already had?  He does not need to
change his subjective probabilities at all in response to the 2nd
measurement outcome, does he?
\evf

\vspace{8pt}

It is not going to be easy, because this in fact is what Schack and I
are actually writing a whole paper about at the moment --- this point
has been the most controversial thing (with the Mermin, Unruh,
Wootters, Spekkens, etc., crowd) that we've said in a while, and it
seems that it's going to require a whole paper to do the point
justice. But I'll still try to give you the skinny of it:
\begin{itemize}
\item
   Q: \ \ He does not need to change his subjective probabilities at all
      in response to the 2nd measurement outcome, does he?
\item
   A: \ \ No he doesn't.
\item
   Q: \ \ Don't you even want to say that the second measurement just
      showed to this observer, as was expected, the value that A
      already had?
\item
   A: \ \ No I don't.
\end{itemize}

The problem is one of the very consistency of the subjective point of
view of quantum states.  The task we set before ourselves is to
completely sever any supposed connections between quantum states and
the actual, existent physical properties of the quantum system.  It
is only from this --- if it can be done, and of course we try to argue
it can be done --- that we get any ``interpretive traction'' (as Chris
Timpson likes to say) for the various problems that plague QM.
[[\ldots]]

This may boil down to a difference between the Rovellian and the
Bayesian/Paulian approach; I'm not clear on that yet. [[\ldots]]
Rovelli relativizes the states to the observer, even the pure states,
and with that --- through the eigenstate-eigenvalue link --- the values
of the observables.  I'm not completely sure what that means in
Rovelli-world yet, however.

I, on the other hand, do know that I would say that a measurement
intervention is always generative of a new fact in the world,
whatever the measurer's quantum state for the system.  If the
measurer's state for the system \emph{happens} to be an eigenstate of the
Hermitian operator describing the measurement intervention, then the
measurer will be confident, \emph{certain} even, of the consequence of the
measurement intervention he is about to perform.  But that \emph{certainty}
is in the sense of Jeffrey and Wittgenstein above --- it is a ``tone of
voice'' of utter confidence.  The world could still, as a point of
principle, smite the measurer down by giving him a consequence that
he predicted to be impossible.  In a traditional development --- with
ties to a correspondence theory of truth --- we would then say, ``Well,
that proves the measurer was wrong with his quantum state assignment.
He was wrong before he ever went through the motions of the
measurement.''  But as you've gathered, I'm not about traditional
developments.  Instead I would say, ``Even from my view there is a
sense in which the measurer's quantum state is \emph{wrong}.  But it is \emph{made
wrong} by the \emph{actual} consequence of the intervention --- it is made
wrong on the fly; its wrongness was not determined beforehand.'' And
that seems to be the main point of contention.
\eq

Particularly this is going to be a key point when I finally come to
the analysis in Section 7 of your paper.
\end{quotation}

\phantomsection
\label{14november2006}
{\bf CAF to Veiko Palge and Bill Demopoulos, 14 November 2006:}
\begin{quotation}
  \noindent The best answer I can give you, I think, is that neither
  of these kinds of structures [sigma-algebras or noncommutative
    generalizations] map onto what I'm thinking.  The main reason for
  this is that I think it is incorrect to think of the process of
  ``quantum measurement'' either 1) as the {\it revelation\/} of a
  property inherent in the system under observation, or 2) as the {\it
    production\/} of such a property in the system.  And without that,
  I don't think there is enough glue to bind the events occurring in
  quantum measurements together into an algebraic structure (say, a
  lattice or a Boolean algebra, or even a partial Boolean algebra,
  where there are Boolean algebras tied together at the edges) --- at
  least not in any useful sense that intrigues me as a
  physicist. [\ldots\!] I think there is a fruitful similarity between
  what [Bill] and I are seeking, even if we ultimately diverge.  It is
  this: In both our views --- and they are the only places I've ever
  seen this style of idea --- even when two measurements share a common
  element (say a given projector $P_i$), there is no implication of a
  common truth value being imposed on $P_i$ across the measurements.
  The reason for this for Bill is that the system's properties are
  bound up with the whole orthogonal family of projectors the
  individual $P_i$ happens to be embedded within.  In contrast, the
  reason for this for me is that I don't think of quantum measurement
  outcomes as signifying properties intrinsic to the system --- they are
  simply consequences of actions for me.  What makes the element $P_i$
  identified across measurements --- for both of us --- is not truth
  value, but that the {\it probabilities\/} for $P_i$ are identical in
  both cases.  What this means particularly from my Bayesian way of
  thinking is that a judgment is being made: I, the agent, am
  identifying {\it this\/} potential outcome of this measurement with
  {\it that\/} potential outcome of that measurement because I judge
  their probabilities equal under all imaginable circumstances.  (See
  Section 4.1 of my \quantph{0205039}.)  It is not that they are
  identified in Nature itself.  Thus, for me, I think, there is no
  good sense in which they lie in the same event space at all.
\end{quotation}

\phantomsection
\label{10january2007}
{\bf CAF to Veiko Palge, 10 January 2007:}
\begin{quotation}
  \noindent Let me start with your first sentence: ``It seems that one
  of the main reasons you reject a well-structured event space is that
  it assumes a realistic interpretation: its elements would correspond
  to intrinsic properties of quantum systems.''  That is the wrong
  direction of reasoning --- though I am probably the cause of this
  misimpression through the restrictive choices of readings I
  recommended to you.  It is not that I reject a well-structured event
  space because it {\it assumes\/} its elements would correspond to
  intrinsic properties of quantum systems, but rather this is the {\it
    result\/} of a thoroughgoing subjective interpretation of
  probabilities within the quantum context.  What cannot be forgotten
  is that quantum-measurement outcomes, by the usual rules, {\it
    determine\/} posterior quantum states.  And those posterior
  quantum states in turn determine further probabilities.

Thus, if one takes the timid, partial move that Itamar and Jeff Bub,
say, advocate --- i.e., simply substituting one or another nonBoolean
algebra for the space of events, and leaving the rest of Bayesian
probability theory seemingly intact --- then one ultimately ends up
re-objectifying what had been initially supposed to be subjective
probabilities.  That is: When I look at the click, and note that it is
value $i$, and value $i$ is rigidly --- or I should say, {\it
  factually} --- associated with the projector $\Pi_i$ in some
nonBoolean algebra, then I have no choice (through L\"uders rule) but
to assign the posterior quantum state $\Pi_i$ to the system.  This
means the new quantum state $\Pi_i$ will be as factual as the click.
And any new probabilities (for the outcomes of further measurements)
determined from this new quantum state $\Pi_i$ will also be factual.

So, the starting point of the reasoning is to {\it assume\/} that
there is a category distinction between probabilities and facts (this
is the subjectivist move of de Finetti and Ramsey).  Adding the
ingredient of the usual rules of quantum mechanics, one derives a
dilemma: If there is a rigid, factual connection between the clicks
$i$ and elements $\Pi_i$ of an algebraic structure, then probabilities
are factual after all.  Holding tight to my assumption of a category
distinction between facts and probabilities, I end up rejecting the
idea that there is a unique, factual mapping between $i$ and $\Pi_i$. [\ldots\!]

[Now let] me return to your paragraph before signing off.  If I read this
question of yours in isolation: ``Can't one just take the elements as
corresponding to clicks and blips in the measurement apparatus?''
Then at one level my answer is, ``Of course; I've never said
otherwise.''  What is at issue here is whether the events --- the clicks
and blips themselves --- fall within the kind of algebraic structure you
speak of, or whether it is something else (something a conceptual
layer above the events) that falls within it.  From the [proto-QBist]
point of view, for a single device with clicks $i$, one agent might
associate the clicks with a set of orthogonal projectors $\Pi_i$, and
another agent might associate them with a set of noncommuting effects
(i.e., POVM elements) $E_i$.  This was the sense in which I meant
there is no stand-alone event space at all: For us, the algebraic
structure of the events (the clicks and blips), their level of
commutivity or noncommutivity and whatnot, is just as subjective as
the quantum state.  The clicks $i$ themselves are objective (in the
sense of not being functions of the subject's beliefs or degrees of
belief), but their association with a particular set of operators
$E_i$ is a subjective judgment.

I hope this completely answers your question now.

But let me extend the discussion a little to try to give you a more
positive vision of what we're up to.  The starting point is the
category distinction between facts and probabilities applied to the
quantum measurement context.  From this we glean that quantum
operations and quantum states are of the same level of subjectivity.
But that is not our ending point.  Because, implicit in everything we
have said there are these autonomous, realistically-interpreted
quantum systems: The agent has to interact with the quantum system to
receive his quantum measurement outcome.  No quantum system, no
measurement outcome.  Thus the [proto-QBist] position is more than a
kind of positivism or operationalism.  The objects of the external
world with which we interact have a certain kind of active power, and
we become aware of the presence of that active power particularly in
the course of quantum measurement.  When we kick on a quantum system
it surprises us with a kick back.

Can we say anything more explicit about the active power?  Can we give
it some mathematical shape?  Yes, I think we can, but that is a
research project.  Still, I think the hints for it are already in
place \ldots\ and indeed they are in the quantum formalism, as we
would expect them to be.  One of the hints is this: The Born
transformation rule.  When we make probability assignments in quantum
mechanics, we are assuming more than de Finettian / Ramseyian
coherence.  We assume that if we set the probabilities for the
outcomes of this measurement this way, then we should set the
probabilities for the outcomes of that measurement that way and the
rule of transformation is a linear one.  This, from the [proto-QBist]
view, is the content of the Born rule (see the last section in the
attached paper).  And it is empirical, contingent: A different world
than the one we live in might have had a different transformation
rule.  So, if we're looking for something beyond personal
probabilities in quantum mechanics, that is a point to take seriously.
It hints of some deep property of our world, and I'd like to know what
that property is. [\ldots\!]

Though [the proto-QBists] banish the algebraic structure of Hilbert
space from having anything to do with a fundamental event space (and
in this way their quantum Bayesianism differs from the cluster of
ideas Pitowsky and Bub are playing with), they do not banish the
algebraic structure from playing any role whatsoever in quantum
mechanics.  It is just that the algebraic structure rears its head at
the conceptual level of coherence rather than in a fundamental event
space.  It is not that potential events are objectively tied to
together in an algebraic way, but that our gambling commitments
(normatively) {\it should be}.  This is another point of contact
between my view and Bill's.
\end{quotation}

\phantomsection
\label{7january2011}
{\bf CAF to Bill Demopoulos, 7 January 2011:}
\begin{quotation}
  \noindent Take a POVM consisting of operators $E_k$.  I.e., these
  are positive semi-definite operators that sum to the identity
  operator.  Thus they might seem like mysterious abstract entities
  (perhaps properties of the system, or properties of the `measuring
  device', or maybe something still weirder, though surely
  agent-independent).  But, via a fiducial SIC POVM in the sky (as I
  talk about in my papers), one can map these operators to a set of
  conditional probabilities $p(i|k)$ in a one-to-one fashion.  That
  is, these operators contain nothing over and above the prescription
  of a probability assignment.  There is nothing to them
  mathematically but that --- nothing is lost by thinking of them in
  these terms, as probability assignments.  Similarly, the
  philosophical point should be this: That there is nothing
  conceptually more than this either.

So, what should one call the ``outcome'' of a quantum measurement?
Well, for the agent it is an experience ``$k$'' \ldots\ but what
meaning does that have for him?  How could he articulate it?  What
does it mean to him?  How will he change his behavior with regard to
it?  The answer is: It is $p(i|k)$.  It is a probability distribution
conditioned on $k$.  The only operational handle the agent has on $k$
is through the assignment he makes, $p(i|k)$.

But all probabilities (for the personalist Bayesian) are subjectively
given (i.e., functions of the agent only, his history and experiences,
not the object).  And thus, I guess, my resistance to part of your way
of thinking about effects.  (Of course, as you should know by now, I
am in serious agreement with you in other ways \ldots\ in fact,
probably with you more than with any other living philosopher.)
\end{quotation}

\end{document}